\documentclass{article}

\usepackage{amsmath}
\usepackage{amssymb}
\usepackage{amsfonts}
\usepackage{mathtools}
\usepackage[left=2.5cm,right=2.5cm]{geometry}

\usepackage[style=numeric]{biblatex}
\usepackage{color,colortbl}
\definecolor{cristianoorange}{rgb}{0.9,0.3,0.}
\definecolor{paoloblue}{rgb}{0.,0,0.8}
\definecolor{piergreen}{rgb}{0.1,0.5,.3}

\usepackage[T1]{fontenc}    
\usepackage[unicode]{hyperref}       
\usepackage{url}            
\usepackage{booktabs}       
\usepackage{amsfonts}       
\usepackage{nicefrac}       
\usepackage{microtype}      
\usepackage{xcolor}         

\usepackage{authblk}

\title{Error-based or target-based? A unifying framework for learning in recurrent spiking networks}

\author[1]{Cristiano Capone\thanks{These authors equally contributed to the article}}
\author[2]{Paolo Muratore$^\ast$}
\author[3]{Pier Stanislao Paolucci}

\affil[1]{INFN, Sezione di Roma, Italy RM 00185}
\affil[2]{SISSA - International School of Advanced Studies, Trieste, Italy}

\begin{document}

\maketitle

\begin{abstract}

Learning in biological or artificial networks means changing the laws governing the network dynamics in order to better behave in a specific situation.
In the field of supervised learning, two complementary approaches stand out: error-based and target-based learning. However, there exists no consensus on which is better suited for which task, and what is the most biologically plausible. Here we propose a comprehensive theoretical framework that includes these two frameworks as special cases. 
This novel theoretical formulation offers major insights into the differences between the two approaches. In particular, we show how target-based naturally emerges from error-based when the number of constraints on the target dynamics, and as a consequence on the internal network dynamics, is comparable to the degrees of freedom of the network.
Moreover, given the experimental evidences on the relevance that spikes have in biological networks, we investigate the role of coding with specific patterns of spikes by introducing a parameter that defines the tolerance to precise spike timing during learning.
Our approach naturally lends itself to Imitation Learning (and Behavioral Cloning in particular) and we apply it to solve relevant closed-loop tasks such as the button-and-food task, and the 2D Bipedal Walker.
We show that a high dimensionality feedback structure is extremely important when it is necessary to solve a task that requires retaining memory for a long time (button-and-food).
On the other hand, we find that coding with specific patterns of spikes enables optimal performances in a motor task (the 2D Bipedal Walker).
Finally, we show that our theoretical formulation suggests protocols to deduce the structure of learning feedback in biological networks.
\end{abstract}

\section{Introduction}

When first confronted with reality, humans learn with high sample efficiency, benefiting from the fabric of society and its abundance of experts in all relevant domains. A conceptually simple and effective strategy for learning in this social context is Imitation Learning. One can conceptualize this learning strategy in the Behavioral Cloning framework, where an agent observes a target, closely optimal behavior (expert demonstration), and progressively improves its mimicking performances by minimizing the differences between its own and the expert's behavior.
Behavioral Cloning can be directly implemented in a supervised learning framework.
In last years competition between two opposite interpretations of supervised learning is emerging: error-based approaches \cite{sacramento2018dendritic,nicola2017supervised,bellec2020,bellec2018long,kreutzer2020natural}, where the error information computed at the environment level is injected into the network and used to improve later performances, and target-based approaches \cite{meulemans2020theoretical,lee2015difference,depasquale2018full,muratore2021target,capone2019sleep,golosio2021thalamo,urbanczik2014learning}, where a target for the internal activity is selected and learned.
In this work, we provide a general framework where these different approaches are reconciled and can be retrieved via a proper definition of the error propagation structure the agent receives from the environment. Target-based and error-based are particular cases of our comprehensive framework. 
This novel formulation, being more general, offers new insights on the importance of the feedback structure for network learning dynamics, a still under-explored degree of freedom.
Moreover, we observe that spike-timing-based neural codes are experimentally suggested to be important in several brain systems \cite{carr1990circuit,johansson2004first,johansson2004first,panzeri2001role,gollisch2008rapid}. This evidence led us to we investigate the role of coding with specific patterns of spikes by introducing a parameter that defines the tolerance to precise spike timing during learning.
Although many studies have approached learning in feedforward  \cite{muratore2021target,memmesheimer2014learning,diehl2015unsupervised, lillicrap2016random,zenke2018superspike, mozafari2019bio} and recurrent spiking networks
\cite{bellec2020,nicola2017supervised,depasquale2018full,ingrosso2019training,capone2019sleep}, a very small number of them successfully faced real world problems and reinforcement learning tasks \cite{bellec2020,traub2021many}.
In this work, we apply our framework to the problem of behavioral cloning in recurrent spiking networks and show how it produces valid solutions for relevant tasks (button-and-food and the 2D Bipedal Walker). 
From a biological point of view, we focus on a tantalizing novel route opened by such a framework: the exploration of what feedback strategy is actually implemented by biological networks and in the different brain areas. 
We propose an experimental measure that can help elucidate the error propagation structure of biological agents, offering an initial step in a potentially fruitful insight-cloning of naturally evolved learning expertise.


\section{Methods}

\subsection{The spiking model}

In our formalism neurons are modeled as real-valued variable $v_j^t \in \mathbb{R}$, where the $j \in \left\{1, \dots, N\right\}$ label identifies the neuron and $t \in \left\{1, \dots, T\right\}$ is a discrete time variable. Each neuron exposes an observable state $s_j^t \in \left\{0, 1 \right\}$, which represents the occurrence of a spike from neuron $j$ at time $t$. 
We then define the following dynamics for our model: 

\begin{align}
&\hat s_i^{t} = \exp{ \left(- \frac{\Delta t}{\tau_s}\right)} \, \hat s_i^{t-1}  +\left( 1 - \exp{ \left(-\frac{\Delta t}{\tau_s} \right) }  \right) \, s_i^{t} \\
&v_i^{t} = \exp{ \left(- \frac{\Delta t}{\tau_m}\right)} \, v_i^{t-1} + \left( 1 - \exp{ \left(-\frac{\Delta t}{\tau_m} \right) }  \right) \, \left( \sum_j  {w}_{ij} \hat s_j^{t-1} + I_i^t + v_\mathrm{rest} \right) - \, w_\mathrm{res} s_i^{t-1}\\
&  s_i^{t+1} = \Theta \left[ v_i^t - v^\mathrm{th}  \right]
\label{p_s}
\end{align}

Where $\Delta t = 1\mathrm{ms}$ is the discrete time-integration step, while $\tau_s= 2\mathrm{ms}$ and $\tau_m= 8\mathrm{ms}$ are respectively the spike-filtering time constant and the temporal membrane constant. Each neuron is a leaky integrator with a recurrent filtered input obtained via a synaptic matrix $ {w} \in \mathbb{R}^{N \times N}$ and an external signal $I^t_i$. $w_\mathrm{res} = -20$ accounts for the reset of the membrane potential after the emission of a spike. $v^\mathrm{th}=0$ and $v_\mathrm{rest}=-4$ are the the threshold and the rest membrane potential.

\subsection{Basics and definitions}

We face the general problem of an agent interacting with an environment with the purpose to solve a specific task.
This is in general formulated in term of an association, at each time $t$, between a state defined by the vector $x^t_h$ and actions defined by the vector $y^t_k$. The agent evaluates its current state and decides an action through a policy $\pi(\{y_k^{t+1}\}|\{x_h^t\})$.
Two possible and opposite strategies to approach the problem to learn an optimal policy are Reinforcement Learning and Imitation Learning.
In the former the agent starts by trial and error and the most successful behaviors are potentiated. In the latter the optimal policy is learned by observing an expert which already knows a solution to the problem.
Behavioral Cloning belongs to the category of Imitation Learning and its scope is to learn to reproduce a set of expert behaviours (actions) $y^{\star \ {t+1}}_k \in\mathbb{R}$, $k = 1,... \, O$ (where $O$ is the output dimension) given a set of states $x^{\star \ t}_h\in\mathbb{R}$, $h = 1,... \, I$ (where $I$ is the input dimension). 
Our approach is to explore the implementation of Behavioral Cloning in recurrent spiking networks.

\subsection{Behavioral Cloning in spiking networks}
In what follows, we assume that the action of the agent at time $t$, $y^{\star \ t}_k $ is evaluated by a recurrent spiking network and can be decoded through a linear readout $y_k ^t = \sum_{i}  {\mathsf{B}}_{ik} \bar s_i ^t$, where $ {\mathsf{B}}_{ik}\in\mathbb{R}$. 
$\bar s_i ^t$ is defined as
\begin{equation}
\bar{s}_i^{t} = \exp{ \left(- \frac{\Delta t}{\tau_\star}\right)} \, \bar{s}_i^{t-1}  + \left( 1 - \exp{ \left(-\frac{\Delta t}{\tau_\star} \right) }  \right) \, s_i^{t}
\label{target_filtering}
\end{equation}
is a temporal filtering of the spikes $ {s}_i ^t$.
To train the network to clone the expert behavior it is necessary to minimize the error:
\begin{equation}
\mathrm{E} = \sum_{t,k} \left( y_{k}^{\star \ t} - y_{k} ^t \right) ^2.
\end{equation}
It is possible to derive the learning rule by differentiating the previous error function (by following the gradient), similarly to what was done in \cite{bellec2020}:
\begin{equation}
    \Delta w_{ij} \propto \frac{d \mathrm{E}}{d w_{ij}} \simeq  \sum_{t} \left[ \sum_k  {\mathsf{B}}_{ik} \left( y_k^{\star t+1} - y_k ^{t+1} \right) \right]  p_i ^t e_j ^t,
    \label{learning_errorbased}
\end{equation}
where we have used $p_i ^t$ for the pseudo-derivative (similarly to \cite{bellec2020}) and reserved $e_j ^t = \frac{\partial v_i^t}{\partial w_{ij}}$ for the spike response function that can be computed iteratively as
\begin{equation}
\frac{\partial v_i^{t+1}}{\partial w_{ij}}  = \exp{ \left(- \frac{\Delta t}{\tau_m}\right)} \, \frac{\partial v_i^{t} }{\partial w_{ij}} +\left( 1 - \exp{ \left(-\frac{\Delta t}{\tau_m} \right) }  \right) \, \hat{s}^{t}_\mathrm{i}.
\label{dv_dynamics}
\end{equation}
In our case the pseudo-derivative, whose purpose is to replace $\frac{\partial f \left( s_i^{t+1} | v_i^t \right)}{\partial v_i^t}$ (since $f(\cdot)$ is non-differentiable, see eq.\eqref{p_s}),  is defined as follows:
\begin{equation}
p_i^t = \frac{e^{v_i^t/\delta v}}{\delta v (e^{v_i^t/\delta v}+1)^2},
\end{equation}
it peaks at $v_i^t = 0$ and $\delta v$ is a parameter defining its width. For the complete derivation we refer to the \textit{supplemental material} (where we also discuss the $\simeq$ in eq. \eqref{learning_errorbased}).

\subsection{Resources}
The code to run the experiments is written in Python 3. 
Simulations were executed on a dual-socket server with eight-core Intel(R) Xeon(R) E5-2620 v4 CPU per socket. The cores are clocked at 2.10GHz with HyperThreading enabled, so that each core can run 2 processes, for a total of 32 processes per server.

\section{Results}

\subsection{Theoretical Results}

\subsubsection{Generalization}

In eq. \eqref{learning_errorbased} we used the expert behavior $y_k^{\star \ t} $ as a target output. 
However, it is possible to imagine that in both biological and artificial systems there are much more constraints, not directly related to the behavior, to be satisfied.
One example is the following: it might be necessary for the network to encode an internal state which is useful to produce the behavior $y_k^{\star \ t}$ and to solve the task (e.g. an internal representation of the position of the agent). The encoding of this information can automatically emerge during training, however to directly suggest it to the network  might significantly facilitate the learning process. This signal is referred as hint in the literature \cite{ingrosso2019training}.
For this reason we introduce a further set of output targets  $q_k^{\star \ t} $, $k = O+1, ... \, D$ and define $Y_k^{\star \ t} $, $ k= 1,... \, D$ as the collection of $y^\star$ and $q^\star$. 
$Y_k^t $ should be decoded from the network activity through a linear readout $Y_k ^t = \sum_{i}  {\mathsf{R}}_{ik}  {s}_i ^t$ and should be as similar as possible to the target. This can be done by minimizing the error $\mathrm{E} = \sum_{k,t} \left( Y_{k}^{\star \ t} - Y_{k} ^t \right) ^2$.
The resulting learning rule is 
\begin{equation}
\Delta w_{ij} = \eta \sum_{t} \left[ \sum_k  {\mathsf{R}}_{ik} \left( Y_k^{\star t+1} - Y_k ^{t+1} \right) \right]  p_i ^t e_j ^t.
\label{learning_errorbased_D}
\end{equation}

\begin{figure}
\centering
\includegraphics[width=\textwidth]{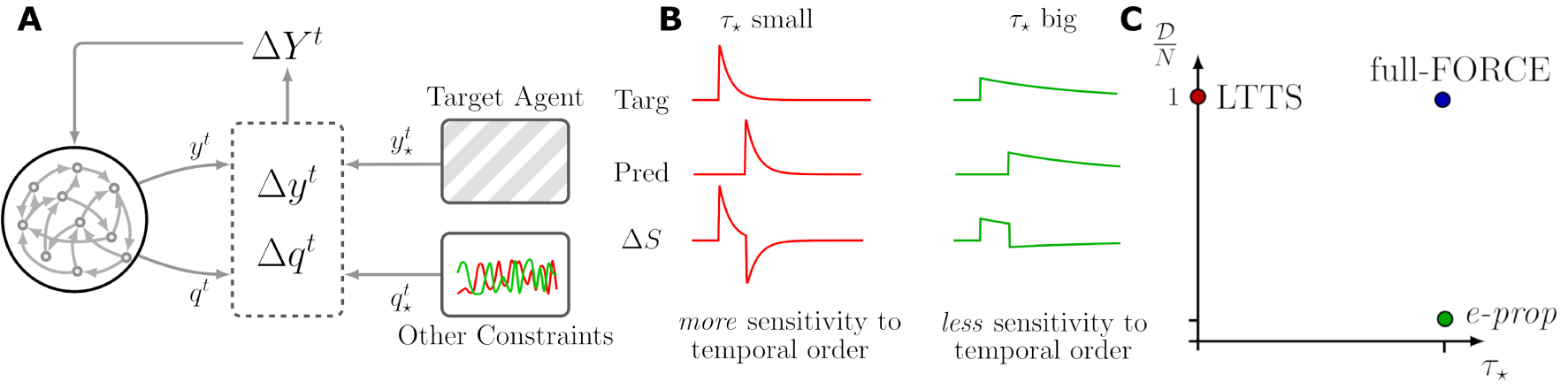}
\caption{\textbf{Framework schematics.} (\textbf{A}) Graphical depiction of a general Behavioural Cloning task. An agent (here a recurrent network) observes the current state-action pair of a target agent and is trained to emulate such behaviour. The model assumes the presence of additional constraints. The total number of independent constraints $D$ defines the rank of the error propagation matrix. (\textbf{B}) Schematics of difference $\tau_\star$, the spikes filtering timescales. A larger $\tau_\star$ is more tolerant on precise spike timing. (\textbf{C}) Schematics of our generalized framework. Changing the $D$ and $\tau_\star$ parameters, it is possible to derive different learning algorithms.}
\label{FIGURE_1}
\end{figure}

\subsubsection{Target-based approach}

The possibility to broadcast specific local errors in biological networks has been debated for a long time \cite{roelfsema2005attention,manchev2020target}. 
On the other hand, the propagation of a target appears to be more coherent with biological observations \cite{knudsen1994supervised,miall1996forward,spratling2002cortical,Larkum2013TINS}.
For this reason we propose an alternative formulations allowing to evaluate target rather than errors \cite{meulemans2020theoretical,manchev2020target}.
This can be easily done by writing the target output as:
\begin{equation}
Y_i^{\star \ t} =   {\mathsf{R}}_{ik} r_k^{\star \ t}.
\end{equation}
Where $r_k^{\star \ t}$ is the target activity of the recurrent network. We observe that if the matrix $\mathsf{R}_{ik}$ is full rank, the internal target can be easily uniquely defined, otherwise it exists a degeneracy in its choice.
Substituting this expression in eq. \eqref{learning_errorbased_D} we obtain
\begin{equation}
\Delta w_{ij} = \eta \sum_t \left[ \sum_k \left(  {\mathsf{R}}^\top  {\mathsf{R}} \right)_{ik} \left( r_k^{\star t+1} - s_k ^{t+1} \right) \right] p_i ^t e_j ^t,
\end{equation}
By inspection, we notice the occurrence of a novel matrix $ {\mathsf{D}} =  {\mathsf{R}}^\top  {\mathsf{R}}$ which acts recurrently on the network, $ {\mathsf{D}} \in \mathbb{R}^{N \times N}$. If one now forgets the origin of this novel matrix, the previous relation can be rewritten in terms of a general square matrix $ {\mathsf{D}} \in \mathbb{R}^{N \times N}$:
\begin{equation}
    \Delta w_{ij} = \eta \sum_t \left[ \sum_k  {\mathsf{D}}_{ik} \left( r_k^{\star t+1} - s_k ^{t+1} \right) \right] p_i ^t e_j ^t.
\end{equation}
The two core new terms are the $r_i^{\star t}$ and the matrix $ {\mathsf{D}}$. The first induces the problem of selecting the optimal network activity, which is tautologically a re-statement of the learning problem. 
The second term, the matrix $ {\mathsf{D}}$ defines the dynamics in the space of the internal network activities $s_k ^t$ during learning. This formulation results similar to the full-FORCE algorithm \cite{depasquale2018full}, which is target-based, but does not impose a specific pattern of spikes for the internal solution.

\subsubsection{Spike coding approximation}

We want now to replace the target internal activity $r_i^{\star \ t}$  with a target sequence of spikes $s_i^{\star\ t}$, in order to approximate the $Y_i^{\star \ t}$ as:
\begin{equation}
Y_i^{\star \ t} \simeq   {\mathsf{R}}_{ik} \bar s_k^{\star \ t}.
\label{spk_approx}
\end{equation}
We stress here the fact that, due to the spikes quantization, the equality cannot be strictly achieved, and eq. \eqref{spk_approx} is an approximation.
One can simply consider $ {s}^{\star t} $ to be the solution of the optimization problem $s_i^{\star t} = \mathrm{argmin}_{s_i^{\star t}} \sum_{kt}| y_k^{\star t}  - \sum_{i}  {\mathsf{B}}_{ki} \bar s_i ^{\star t} |$.
The optimal encoding for a continuous trajectory through a pattern of spikes has been broadly discussed in \cite{brendel2020learning}.
However, the pattern $ {s}^{\star t}$ might describe an impossible dynamics (for example activity that follows periods of complete network silence).
For this reason here we take a different choice. The $s_i^{\star t}$ is the pattern of spikes expressed by the untrained network when the target output $Y_i^{\star t}$ is randomly projected as an input (similarly to \cite{depasquale2018full,muratore2021target}). It has been demonstrated that this choice allows for fast convergence and encodes detailed information about the target output.
With these additional considerations, we can now rewrite our expression for the weight update in terms of the network activity:
\begin{equation}
    \Delta w_{ij} = \eta \sum_t \left[ \sum_k  {\mathsf{D}}_{ik} \left( \bar s_k^{\star t+1} - \bar s_k ^{t+1} \right) \right] p_i ^t e_j ^t.
    \label{rank_rule}
\end{equation}
In this way a specific pattern of spikes is directly suggested to the network as the internal solution of the task.
We observe that when  $ {\mathsf{R}}$ is random and full rank,  $ {\mathsf{D}}$ it is almost diagonal 
and the training of recurrent weights reduces to learning a specific pattern of spikes \cite{pfister2006optimal,rezende2014stochastic,gardner2016supervised,brea2013matching,capone2018spontaneous}.
In this limit the model LTTS \cite{muratore2021target} is recovered (see Fig.\ref{FIGURE_1}C), with the only difference of the presence of the pseudo-derivative.
We interpret the parameter $\tau_{\star}$ (the time scale of the spike filtering, see eq. \eqref{target_filtering}) as the tolerance to spike timing. In Fig.\ref{FIGURE_1}B we show in a sketch that, for the same spike displacement between the internal and the target activity, the error is higher when the $\tau_{\star}$ is lower.

\subsection{Numerical Results}

\subsubsection{Dimensionality of the solution space}

\begin{figure}
\centering
\includegraphics[width=\textwidth]{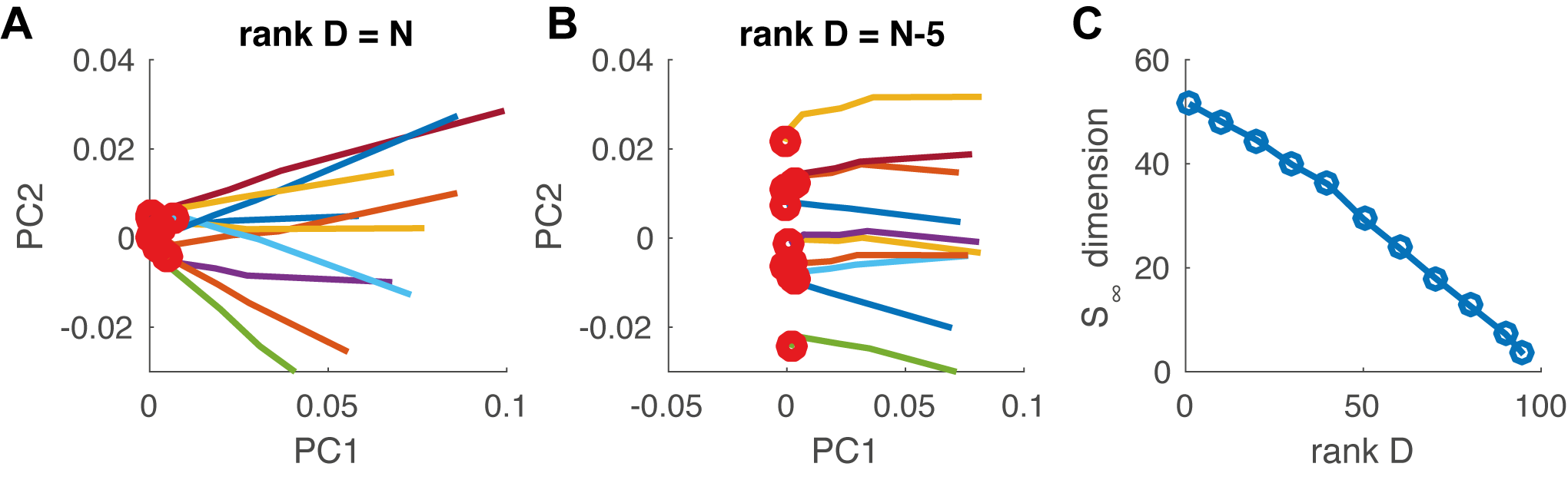}
\caption{\textbf{Error propagation and dimensionality of the internal solution.} 
(\textbf{A}). Dynamics along training epochs of the $\Delta S_i = \sum_t \left| s_i^{\star\ t} - s_i^t \right|$ in the first two principal components for different repetition of the training with variable initial conditions. The error propagation matrix has maximum rank ($\mathcal{D} = N$, target-based limit). 
(\textbf{B}). Same as in (\textbf{A}), but with an error propagation matrix with rank $\mathcal{D} = N - 5$. 
(\textbf{C}). Dimensionality of the solution space $\mathcal{S}_{\infty}$ as a function of the rank $\mathcal{D}$ of the error propagation matrix.
}
\label{FIGURE_2}
\end{figure}

The learning formulation of eq. \eqref{rank_rule} offers a major insights on the role played by the feedback matrix $ {\mathsf{D}}_{ik}$. Consider the learning problem (with fixed input and target output) where the synaptic matrix $w_{ij}$ is refined to minimize the output error (by converging to the proper internal dynamics). 
The learning dynamics can be easily pictured as a trajectory 
where a single point is a complete history of the network activity $ {s}_n =  \left\{s_i^t: i= 1,...N; t = 1,...T  \right\}$.
Upon initialization, a network is located at a point $ {s}_0$ marking its untrained spontaneous dynamics. The following point $ {s}_1$ is the activity produced by the network after applying the learning rule defined in eq. \eqref{rank_rule}, and so on.
By inspecting eq. \eqref{rank_rule} one notes that a sufficient condition for halting the learning is $|\sum_i  {\mathsf{D}}_{hi} \left(\bar{s}_i^{\star \ t} - \bar{s_i}^t \right)|<\epsilon$, where $\epsilon$ is an arbitrary small positive number. If $\epsilon$ is small enough it is possible to write:
\begin{equation}
\sum_i  {\mathsf{D}}_{hi} \left(\bar{s}_i^{\star \ t} - \bar{s_i}^t \right) \simeq 0.
\label{s_dyn_conv}
\end{equation}
In the limit of a full-rank $ {\mathsf{D}}$ matrix (example: the LTTS limit where $ {\mathsf{D}}$ is diagonal) the only solution to eq. \eqref{s_dyn_conv} is $\bar{ {s}}^{t} \simeq \bar{ {s}}^{\star t}$ and the learning halts only when target $\bar{ {s}}^{\star \ t}$ is cloned. 
When the rank is lower the solution to eq. \eqref{s_dyn_conv} is not unique, and the dimensionality of possible solutions is defined by the kernel of the matrix $ {\mathsf{D}}$ (the collection of vectors ${\lambda}$ such that $ {\mathsf{D}} {\lambda} = 0$). We have: $\mathrm{dim}\  \mathrm{Ker}\  {\mathsf{D}} = N - \mathrm{rank}\  {\mathsf{D}} = N - \mathcal{D}$. 
We run a numerical experiment in order to confirm our theoretical predictions. We used equation \eqref{rank_rule} to store and recall a 3D continuous trajectory $y_k^{\star \ t}$ ($k = 1,..3$, $t = 1,..T$, $T=100$) in a network of $N = 100$ neurons. $y_k^{\star \ t}$ is a temporal pattern composed of $3$ independent continuous signals.
Each target signal is specified as the superposition of the four frequencies $f \in \left\{1, 2, 3, 5 \right\}$ Hz with uniformly extracted random amplitude $A \in \left[0.5, 2.0 \right]$ and phase $\phi \in \left[0, 2 \pi \right]$.
We repeated the experiment for different values of the rank $\mathcal{D}$.
The matrix $ {\mathsf{R}}_{ik} = \frac{\delta_{ik}}{\sqrt{\mathcal{D}}}$, $i = 1,...N $, $k = 1,... \mathcal{D}$, where $\delta_{ik}$ is the Kronecker delta (the analysis for the case $ {\mathsf{R}}_{ik}$ random provides analogous results and is reported in the \textit{supplemental material}).
When the rank is $N$, different replicas of the learning (different initializations of recurrent weights) converge almost to the same internal dynamics $s_i^t$. This is reported in Fig.\ref{FIGURE_2}A where a single trajectory represents the first 2 principal components (PC) of the vector $\sum_t |s_i^{\star \ t}-s_i^t|$. The convergence to the point $(0,0)$ represents the convergence of the dynamics to $s_i^{\star\ t}$. When the rank is lower ($\mathcal{D} = N-5$, see Fig.\ref{FIGURE_2}B) different realizations of the learning converge to different points, distributed on an line in the PC space.
This can be generalized by investigating the dimension of the convergence space as a function of the rank. The dimension of the vector $s_i^{\star \ t}-s_i^t$ evaluated in a the trained network is estimated as $d = \frac{1}{\sum_k \lambda_k^2}$, where $\lambda_k$ are the principal component variances normalized to one ($\sum_k \lambda_k$ = 1).
We found a monotonic relation between the dimension of the convergence space and the rank (see Fig.\ref{FIGURE_2}C, more information on the PC analysis and the estimation of the dimensionality in the \textit{supplemental material}).
This observation confirms that when the rank is very high, the solution is strongly constrained, while when the rank becomes lower, the internal solution is free to move in a subspace of possible solutions.
We suggest that this measure can be used in biological data to estimate the dimensionality of the learning constraints in biological neural network from the dimensionality of the solution space.

\subsubsection{Tolerance to spike timing}

\begin{figure}
\centering
\includegraphics[width=\textwidth]{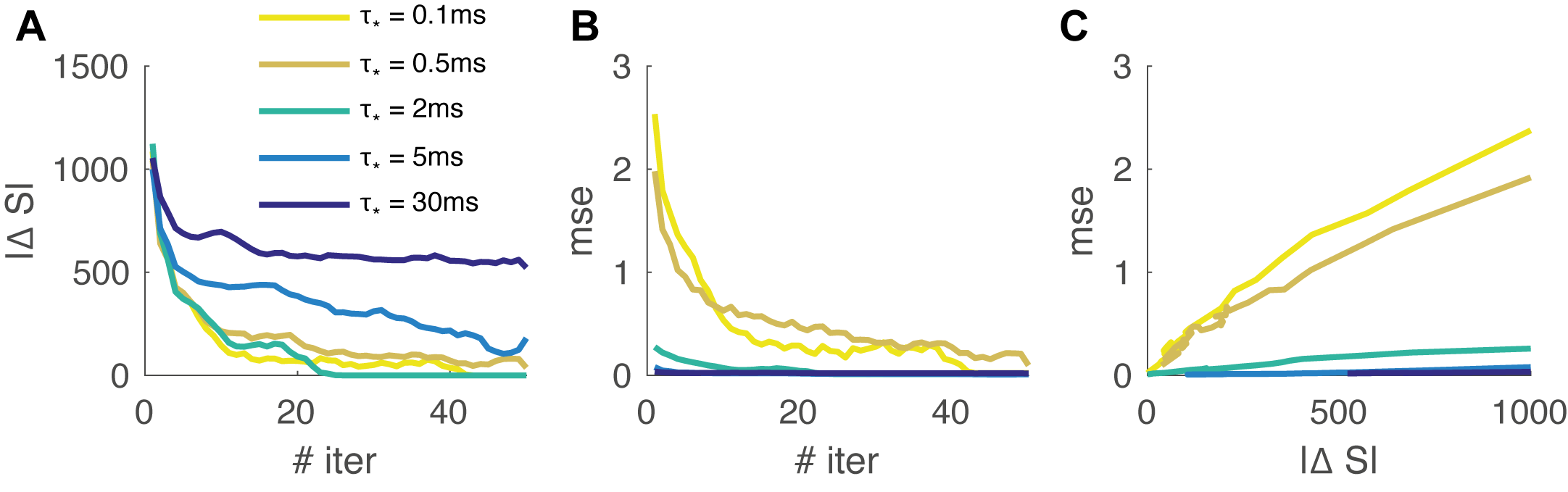}    
\caption{\textbf{Target-based learning for different time-scales.} 
(\textbf{A}). Color-coded the error on the spike sequence $\Delta S = \sum_{it}|s_i^{\star \ t}-s_i^t|$ as a function of the number of iterations for different $\tau_{\star}$.
(\textbf{B}). Color-coded the $\mathrm{mse} = \sum_{kt}(y_k^{\star \ t} - y_k^t)^2$ as a function of the number of iterations for different $\tau_{\star}$.
(\textbf{C}). Scatter plot of $\mathrm{mse}$ vs $\Delta S $ for different values of  $\tau_{\star}$.
}
\label{FIGURE_3}
\end{figure}

As we discussed above the  $\tau_{\star}$ can be interpreted as the tolerance to precise spike timing. To investigate the role of this parameter, we considered the same store and recall task of a 3D trajectory described in the previous section ($N = 100$, $T = 100$).
We set the maximum rank ($\mathcal{D} = N$) for this experiment. In Fig.\ref{FIGURE_3}A we report the spike error $\Delta S = \sum_{it} \left| s_i^{\star \ t} - s_i^t \right|$ as a function of the iteration number for different values of the parameter $\tau_{\star}$. Only for the lower values of $\tau_{\star}$ the algorithm converges exactly to the spike pattern $s_i^{\star \ t}$. In Fig.\ref{FIGURE_3}B we report the $\mathrm{mse} = \sum_{kt}(y_k^{\star \ t} - y_k^t)^2$ as a function of the iteration numbers and the parameter $\tau_{\star}$.
In Fig.\ref{FIGURE_3}C we show the $\mathrm{mse}$ as a function of $\Delta S$ for different values of $\tau_{\star}$. Lower $\tau_{\star}$ values are characterized by a higher slope meaning that a change in the spike pattern expressed by the network strongly affects the error on the output $y_k^t$. This suggests a low tolerance to precise spike timing in the generated output when the parameter $\tau_{\star}$ is low. The consequence of this effect in a behavioral task is investigated below (section 2D Bipedal Walker).

\subsection{Application to closed-loop tasks}

\subsubsection{Button-and-food task}

\begin{figure}
\centering
\includegraphics[width=\textwidth]{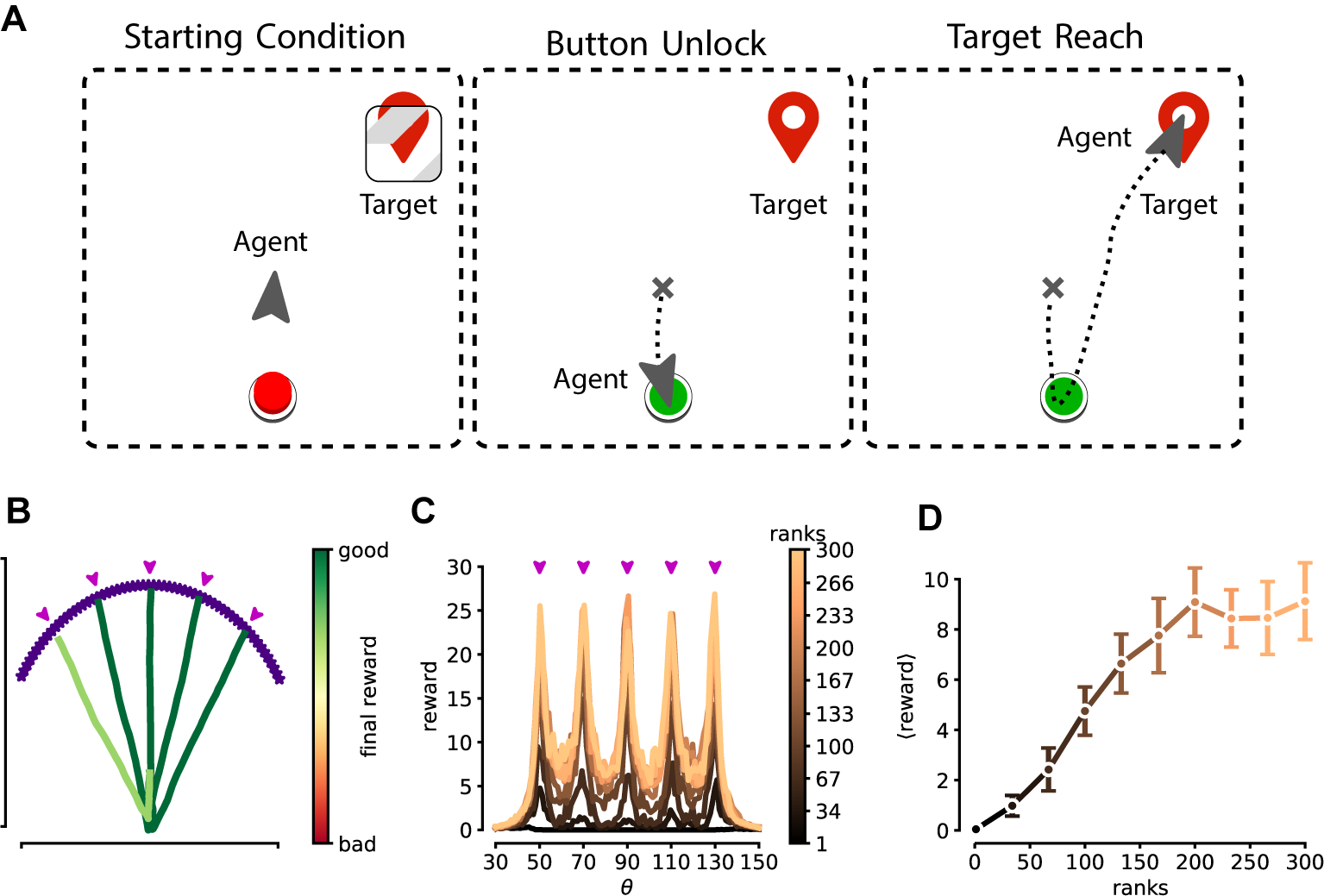}
\caption{\textbf{Button-and-food task.} (\textbf{A}) Sketch of the task. An agent start at the center of the environment domain (left) and is asked to reach a target. The target is initially "locked". The agent must unlock the target by pushing a button (middle) placed behind and then reach for the target (right). (\textbf{B}) Example trajectories produced by a trained agent for different target locations. Purple arrows depict the observed expert behaviours. (\textbf{C}) Final reward obtained by a trained agents as a function of the target position (measured by the angle $\theta$ with a fixed radius of $r = 0.7$ as measured from the agent starting position). Continuous lines are average values, while error bars are standard deviation for $10$ repetitions. Colors codes for different ranks in the error propagation matrix. (\textbf{D}) Average reward as a function of the rank.}
\label{FIGURE_4}
\end{figure}

To investigate the effect of the rank $\mathcal{D}$ of feedback matrix, we design a button-and-food task (see Fig.\ref{FIGURE_4}A for a graphical representation), which requires for a precise trajectory and to retain the memory of the past states. In this task, the agent starts at the center of the scene, which features also a button and an initially locked target (the food). The agent task is to first push the button so to unlock the food and then reach for it. We stress that to change its spatial target from the button to the food, the agents has to remember that it already pressed the button (the button state is not provided as an input to the network during the task).
In our experiment we kept the position of the button (expressed in polar coordinates) fixed at $r_\mathrm{btn} = 0.2$, $\theta_\mathrm{btn} = -90^\circ$ for all conditions, while food position had $r_\mathrm{targ} = 0.7$ and variable $\theta_\mathrm{targ} \in \left[30^\circ, 150^\circ \right]$. 
The agent learns via observations of a collection of experts behaviours, which we indicate via the food positions $\left\{ \theta_\mathrm{targ}^\star \right\}$. The expert behavior is a trajectory which reaches the button and then the food in straight lines ($T = 80$). 
The network receives as input ( $I = 80$ input units) the vertical and horizontal differences  of both the button's and food's positions with respect to agent location ($\Delta^t =\{\Delta x_b^t, \Delta y_b^t,\Delta x_f^t, \Delta y_f^t\}$ respectively). These quantities are encoded through a set of tuning curves. Each of the $\Delta_i$ values are encoded by 20 input units with different Gaussian activation functions.
Agent output is the velocity vector $v_{x, y}$ ($O = 2$ output units). We used $\eta = \eta_\mathrm{RO} = 0.01$ (with Adam optimizer), moreover $\tau_\mathrm{RO} = 10\mathrm{ms}$.
Agent performances are measured with the inverse of final agent-food distance for unlocked food $r_\mathrm{unlocked} = d \left(p_\mathrm{agent}, p_\mathrm{target} \right)^{-1}$, and kept fixed at $r_\mathrm{locked} = 1 / d_\mathrm{max}$ (with $d_\mathrm{max} = 5$) for the locked condition. We repeated training for different values of the rank of the feedback matrix $ {\mathsf{D}}$, computed from $ {\mathsf{R}}_{ik} = \frac{\delta_{ik}}{\sqrt{\mathcal{D}}}$ (with $  \delta_{ik}$ the Kronecker delta, the analysis for the case $ {\mathsf{R}}_{ik}$ random provides analogous results and is reported in the \textit{supplemental material}), in a network of $N = 300$ neurons, and compared the overall performances (more information in the \textit{supplemental material}). Results for such experiment are reported in Fig.\ref{FIGURE_4}B-C. Fig.\ref{FIGURE_4}B, we report the agent training trajectories, color-coded for the final reward. Indeed all the training conditions ($\theta^\star_\mathrm{targ} \in \left\{50, 70, 90, 110, 130 \right\}$) show good convergence. In Fig.\ref{FIGURE_5}C the final reward is reported as a function of the target angle $\theta_\mathrm{targ}$ for different ranks (purple arrows indicate the training conditions). As expected, the reward is maximum concurrently to the training condition. Moreover, it can be readily seen how high-rank feedback structures allows for superior performances for this task.

\subsubsection{2D Bipedal Walker}

\begin{figure}
\centering
\includegraphics[width=\textwidth]{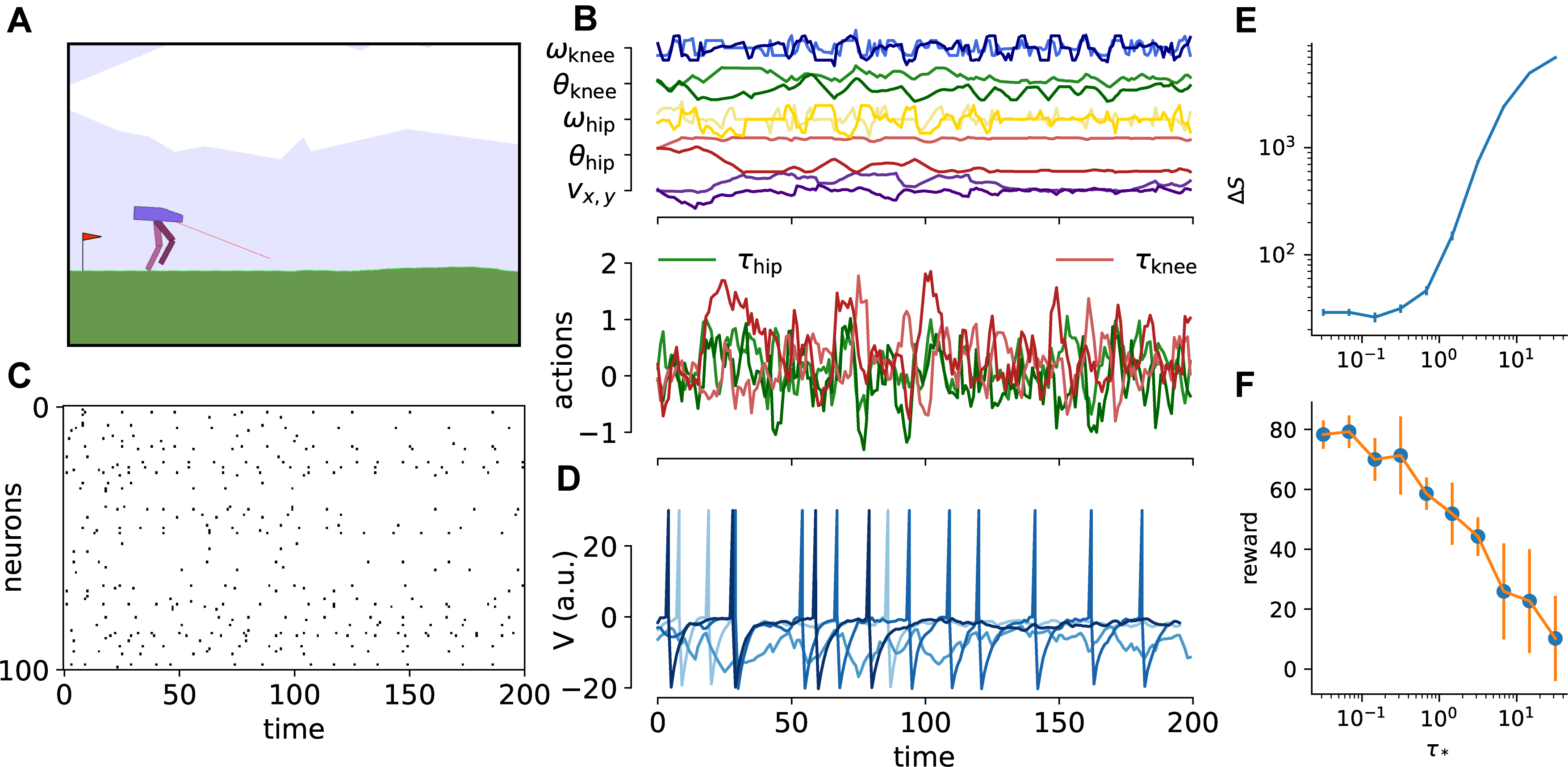}
\caption{\textbf{2D Bipedal Walker.}
(\textbf{A}). Representation of the 2D Bipedal Walker environment. The task is to successfully control the bipedal locomotion of the agent, reward is measured as the travelled distance across the horizontal direction. 
(\textbf{B})  (Top) Temporal dynamics of the core input state variable. $v_{x, y}$ is the velocity direction in the bi-dimensional plane. $\omega_{\mathrm{knee}, \mathrm{hip}}$ are the angular velocity of the two leg joints, while $\theta_{\mathrm{knee}, \mathrm{hip}}$ are the corresponding angles. Each angular variable has two components which corresponds to the two legs of the agent, which are depicted via different shades of the same color. (Bottom) Temporal dynamics of the action vector. The network controls the agent via the torque $\tau_{\mathrm{hip}, \mathrm{knee}}$ that is applied to the two legs. Controls for left and right legs are depicted via different shades of the same color.
(\textbf{C}). Rasterplot of the activity of a random sample of $100$ neurons across $200$ time unit of a task episode.  
(\textbf{D}).  Temporal dynamics of the membrane potential of four example units. 
(\textbf{E}). Spike error $\Delta S$ as a function of the $\tau_{\star}$.
(\textbf{F}). Average reward as a function of the $\tau_{\star}$.
} 
\label{FIGURE_5}
\end{figure}

We benchmarked our behavioral cloning learning protocol on the standard task the 2D Bipedal Walker, provided through the OpenAI gym (\url{https://gym.openai.com} \cite{1606.01540}, MIT License). 
The environment and the task are sketched in Fig.\ref{FIGURE_5}A: a bipedal agent has to learn to walk and to travel as long a distance as possible.
The expert behavior is obtained by training a standard feed-forward network with PPO (proximal policy approximation \cite{schulman2017proximal}, in  particular we used the code provided in \cite{pytorch_minimal_ppo}, MIT License).
The sequence of states-actions is collected in the vectors $ {y}^{\star \ t}_k $, $k = 1,... \, O$, $ {x}^{\star \ t}_h$, $h = 1,... I$, $t = 1,...T$, with $T = 400$, $O = 4$, $I = 15$ (see In Fig.\ref{FIGURE_5}C for an example of the states-actions trajectories). The average reward performed by the expert is $\langle r \rangle_{exp} \simeq 180$ while a random agent achieves $\langle r \rangle_{rnd} \simeq -120$.
We performed behavioral cloning by using the learning rule in eq. \eqref{rank_rule} in a network of $N = 500$ neurons. We chose the maximum rank ($\mathcal{D} = N$) and evaluate the performances for different values of $\tau_{\star}$  (more information in the \textit{supplemental material}).
In Fig.\ref{FIGURE_5}B-C it is report the rastergram for $100$ random neurons and the dynamics of the membrane potential for $3$ random neurons during a task episode.
For each value of $\tau_{\star}$ we performed $10$ independent realizations of the experiment. For each realization the $s_i^{\star \ t}$ is computed, and the recurrent weights are trained by using eq. \eqref{rank_rule}.  The optimization is performed using gradient ascent and a learning rate $\eta = 1.0$.
In Fig.\ref{FIGURE_5}D we report the spike error $\Delta S = \sum_{it} \left| s_i^{\star\ t} - s_i^t \right|$ at the end of the training. The internal dynamics $s_i^t$ almost perfectly reproduces the target pattern of spikes $s_i^{\star \ t}$ for $\tau_{\star} < 0.5 \mathrm{ms}$, while the error increases for larger values. 
The readout time-scale is fixed to $\tau_\mathrm{RO} = 5\mathrm{ms}$ while the readout weights are initialized to zero and the learning rate is set to $\eta_\mathrm{RO} = 0.01$.
Every $75$ training iterations of the readout we test the network and evaluate the average reward $\langle r \rangle$ over $50$ repetitions of the task. We then evaluate the average over the $10$ realizations of the maximum $\langle r \rangle$ obtained for each realization.
In Fig.\ref{FIGURE_5}F it is reported the average of the maximum reward as a function of $\tau_{\star}$. The decreasing monotonic trend suggests that learning with specific pattern of spikes ($  \tau_{\star} \rightarrow 0$) enables for optimal performances in this walking task.
We stress that in this experiment we used a clumped version of the learning rule. In other words we substituted $s_i^{\star \ t}$ to $s_i^{t}$ in the evaluation of $\frac{\partial v_i}{\partial w_{ij}}$ in eq.\eqref{dv_dynamics}. This choice, which is only possible when the maximum rank is considered ($\mathcal{D} = N$), allows for faster convergence and better performances. The results for the non-clumped version of the learning rule are reported in the \textit{supplemental material}.

\section{Discussion}
In this work, we introduced a general framework for supervised learning in recurrent spiking networks, with two main parameters, the rank of the feedback error propagation $\mathcal{D}$ and the tolerance to precise spike timing $\tau_{\star}$ (see Fig.\ref{FIGURE_1}C). We argue that many proposed learning rules can be seen as specific cases of our general framework (e-prop, LTTS, full-FORCE).
In particular, the generalization on the rank of the feedback matrix allowed us to understand the target-based approaches as emerging from error-based ones when the number of independent constraints is high.
Moreover, we understood that different $\mathcal{D}$ values lead to different dimensionality of the solution space. If we see the learning as a trajectory in the space of internal dynamics, when the rank $\mathcal{D}$ is maximum, every training converges to the same point in this space. On the other hand, when the $\mathcal{D}$ is lower, the solution is not unique, and the possible solutions are distributed in a subspace whose dimensionality is inversely proportional to the rank of the feedback matrix.
We suggest that this finding can be used to produce experimental observable to deduce the actual structure of error propagation in the different regions of the brain.
On a technological level, our approach offers a strategy to clone on a (spiking) chip an expert behavior either previously learned via standard reinforcement learning algorithms or acquired from a human agent.
Our formalism can be directly applied to train an agent to solve closed-loop tasks through a behavioral cloning approach. This allowed solving tasks that are relevant in the reinforcement learning framework by using a recurrent spiking network, a problem that has been faced successfully only by a very small number of studies \cite{bellec2020}. Moreover, our general framework,  encompassing different learning formulations, allowed us to investigate what learning method is optimal to solve a specific task.
We demonstrated that a high number of constraints can be exploited
to obtain better performances in a task in which it was required to retain a memory of the internal state for a long time (the state of the button in the button-and-food task).
On the other hand, we found that a typical motor task (the 2D Bipedal Walker) strongly benefits from precise timing coding, which is probably due to the necessity to master fine movement controls to achieve optimal performances. In this case, a high rank in the error propagation matrix is not really relevant.
From the biological point of view, we conjecture that different brain areas might be located in different positions in the plane presented in Fig.\ref{FIGURE_1}C.
\subsection{Limitations of the study}
We chose relevant but very simple tasks in order to test the performances of our model and understand its properties. However, it is very important to demonstrate if this approach can be successfully applied to more complex tasks, e.g. requiring both long-term memory and fine motor skills. It would be of interest to measure what are the optimal values for both the rank of feedback matrix and $\tau_\star$in a more demanding task.
Finally, we suggested that our framework allows for inferring the error propagation structure. 
However, our measure requires knowing the target internal dynamics which is not available in experimental recordings. We plan to develop a variant of this measure that doesn't require such an observable.
Moreover, we observe that the measure we proposed is indirect since it is necessary to estimate the dimensionality of the solution space first and then deduce the dimensionality of the learning constraints. Future development of the theory might be to formulate a method that directly infers from the data the laws of the dynamics in the solution space induced by learning.

\section*{Acknowledgement}
This work has been supported by the European Union Horizon 2020 Research and Innovation program under the FET Flagship Human Brain Project (grant agreement SGA3 n. 945539 and grant agreement SGA2 n. 785907) and by the INFN APE Parallel/Distributed Computing laboratory.

\printbibliography

\end{document}


\maketitle

\appendix

\section{Model and learning rule}

\subsection{The spiking model}

In our formalism neurons are modeled as real-valued variable $v_j^t \in \mathbb{R}$, where the $j \in \left\{1, \dots, N\right\}$ label identifies the neuron and $t \in \left\{1, \dots, T\right\}$ is a discrete time variable. Each neuron exposes an observable state $s_j^t \in \left\{0, 1 \right\}$, which represents the occurrence of a spike from neuron $j$ at time $t$. 
We then define the following dynamics for our model: 

\begin{align}
&\hat s_i^{t} = \exp{ \left(- \frac{\Delta t}{\tau_s}\right)} \, \hat s_i^{t-1}  +\left( 1 - \exp{ \left(-\frac{\Delta t}{\tau_s} \right) }  \right) \, s_i^{t} \\
&v_i^{t} = \exp{ \left(- \frac{\Delta t}{\tau_m}\right)} \, v_i^{t-1} + \left( 1 - \exp{ \left(-\frac{\Delta t}{\tau_m} \right) }  \right) \, \left( \sum_j  {w}_{ij} \hat s_j^{t-1} + I_i^t + v_\mathrm{rest} \right) - \, w_\mathrm{res} s_i^{t-1}\\
&  s_i^{t+1} = \Theta \left[ v_i^t - v^\mathrm{th}  \right]
\label{p_s}
\end{align}


$\Delta t$ is the discrete time-integration step, while $\tau_s$ and $\tau_m$ are respectively the spike-filtering time constant and the temporal membrane constant. Each neuron is a leaky integrator with a recurrent filtered input obtained via a synaptic matrix $ {w} \in \mathbb{R}^{N \times N}$ and an external signal $I^t_i$. $w_\mathrm{res} = -20$ accounts for the reset of the membrane potential after the emission of a spike. $v^\mathrm{th}=0$ and $v_\mathrm{rest}=-4$ are the the threshold and the rest membrane potential.

\subsection{Error-based learning rule, complete derivation}

 We derive here the expression for the synaptic update (eq. (6) in the main text) that is obtained in the error-based framework for the minimization of an output error $\mathrm{E}$. We assume a regression problem where the error is computed as:
 
 \begin{equation}
     \mathrm{E} = \sum_{t, k} \left( y_k^{\star t} - y_k^t \right)^2, 
\label{EQ:error_definition}
 \end{equation}
 
 moreover we assume the system output $y_k^t \in \mathbb{R}$ to be a linear readout (via readout real matrix $\mathsf{B}_{ik}$) of the low-pass filtered network activity $\bar{s}_i^t$:
 
 \begin{equation}
     y_k^t = \sum_i \mathsf{B}_{ik} \bar{s}_i^t
\label{EQ:output_definition}
 \end{equation}
 
 where $\bar{s}_i^t$ can be evaluated iteratively as:
 
\begin{equation}
\bar{s}_i^{t} = \beta_\mathrm{RO} \, \bar{s}_i^{t-1}  + \left( 1 - \beta_\mathrm{RO}   \right) \, s_i^{t}.
\label{target_filtering}
\end{equation}

with $\beta_\mathrm{RO} =\exp{ \left(- \Delta t / \tau_\mathrm{RO} \right)}$. The resulting formulation for the synaptic update $\Delta w_{ij}$ is obtained by imposing it to be proportional to the negative error gradient, with the proportionality factor $\eta$ representing the learning rate of the system. Following a classical factorization of the error gradient computation in recurrent network, one start by noticing how the time-unrolled network has a feed-forward structure with layers indexed by the time variable $t$ and shared weights $w_{ij}^t = w_{ij} \ \forall t$. The gradient for a specific time-layer can be expressed as (we use the superscript $w^{(t)}_{ij}$ to denote the formal dependence of the synaptic matrix on the layer of time-unrolled, however being the network recurrent, we will omit this trivial index in subsequent expressions): 
 
 \begin{equation}
     \frac{d \mathrm{E}}{d w_{ij}^{(t)}} = \frac{d \mathrm{E}}{d v_i^t} \frac{\partial v_i^t}{\partial w_{ij}^{(t)}}.
 \end{equation}
 
 The total gradient is obtained by summing the contributions from all the time-layers, yielding the following expression for the error-gradient synaptic update:
 
 \begin{equation}
     \Delta w_{ij} = -\eta \frac{d \mathrm{E}}{d w_{ij}} = - \sum_t \frac{d \mathrm{E}}{d v_i^t} \frac{\partial v_i^t}{\partial w_{ij}}.
\label{EQ:error_factorization}
 \end{equation}
 
Following \cite{bellec2020}, we rewrite the error total derivative by collecting all the terms that can be computed locally. The core issue is the fact that the error $\mathrm{E} = \mathrm{E} \left(s^1, s^2, \dots, s^T \right)$ (with $s^t = \left\{ s_i^t \right\}, \forall i$), is a function of the complete network activity, so the influence of a spike $s_i^t$ on the subsequent network development should be backtracked in the computation of the total derivative. We aim for a recursive rewriting by noting that:

\begin{equation}
    \frac{d \mathrm{E}}{d v_i^t} = \frac{d \mathrm{E}}{d s_i^{t+1}} \frac{\partial s_i^{t+1}}{\partial v_i^t} + \frac{d \mathrm{E}}{d v_i^{t+1}} \frac{\partial v_i^{t+1}}{\partial v_i^t}.
\end{equation}

Indeed the second term is suited for an analogous manipulation. The recursive chain terminates for $v_i^{T+1}$, having the error no dependence for such variable. Applying this strategy yields the following expression for the error gradient:

\begin{equation}
    \frac{d \mathrm{E}}{d w_{ij}} = \sum_{t} \left[ \frac{d \mathrm{E}}{d s_i^{t+1}} \frac{\partial s_i^{t+1}}{\partial v_i^t} + \left( \frac{d \mathrm{E}}{d s_i^{t+2}} \frac{\partial s_i^{t+2}}{\partial v_i^{t+1}} + \left[ \dots \right] \frac{\partial v_i^{t+2}}{\partial v_i^{t+1}} \right) \frac{\partial v_i^{t+1}}{\partial v_i^t} \right] \frac{\partial v_i^t}{\partial w_{ij}}.
\end{equation}

Collecting all the terms of the form $\partial v_i^{t+1} / \partial v_i^t$  yields the following compact expression for the gradient:

\begin{equation}
    \frac{d \mathrm{E}}{d w_{ij}} = \sum_{\tau} \sum_{t > \tau} \frac{d \mathrm{E}}{d s_i^{t + 1}} \frac{\partial s_i^{t + 1}}{\partial v_i^{t}} \left[\frac{\partial v_i^{t}}{\partial v_i^{t-1}} \dots \frac{\partial v_i^{\tau +1}}{\partial v_i^{\tau}} \right] \frac{\partial v_i^{\tau}}{\partial w_{ij}}.
\end{equation}

The current form still bares the issue of require computation of future event (terms indexed by $t > \tau$). However this problem is only apparent (see \cite{bellec2020}), as it can be solved by exchanging the summation indices and rewriting the former expression in a form that, at each time $t$, only involves past events (thus being physically plausible):

\begin{equation}
    \frac{d \mathrm{E}}{d w_{ij}} = \sum_t \frac{d \mathrm{E}}{d s_i^{t+1}} \frac{\partial s_i^{t+1}}{\partial v_i^t} \sum_{\tau < t} \left( \frac{\partial v_i^{t}}{\partial v_i^{t-1}} \dots \frac{\partial v_i^{\tau +1}}{\partial v_i^{\tau}} \right) \frac{\partial v_i^{\tau}}{\partial w_{ij}}.
\end{equation}

If we recognize in the last term  $\partial v_i^t / \partial w_{ij} = \hat s_j^t$ 
and note how $\partial v_i^t / \partial v_i^{t-1} = \beta_m$, where $\beta_m = \exp \left(-\Delta t / \tau_m \right)$, the second summation in the gradient is recognized as a low-pass filter of $\hat{s}_j^t$, which yields the spike response function, namely:

\begin{equation}
    e_j^t = \sum_{\tau < t} \beta_m^{t - \tau} \hat s^t.
\end{equation}

We stress that this formulation is equivalent to eq. (7) of the main text. Inspecting the term $\partial s_i^{t+1} / \partial v_i^t$ we note how, according to eq. (3) of the main text, the spike $s_i^t$ depends in a non-differentiable way on the neuron voltage $v_i^t$ (via $\Theta \left[ v_i^t - v^\mathrm{th} \right]$). This is a fundamental characteristic of spike-based system, which is usually dealt with the introduction of a custom, non-linear pseudo-derivative $p_i^t$ (see eq. (8) of main text). With the previous substitution in place the gradient reads:

\begin{equation}
    \frac{d \mathrm{E}}{d w_{ij}} = \sum_t \frac{d \mathrm{E}}{d s_i^{t+1}} p_i^t e_j^t.
\label{EQ:exact_error_gradient}
\end{equation}

Up until this point all the manipulation have yielded an exact expression. However, the computation of the total derivative of the error with respect to the neuron spike still needs to be accounted for. Again we face the problem of the cascading influence of the term $s_i^t$ on the entire future network activity. In \cite{bellec2020} the following approximation is introduced:

 \begin{equation}
     \frac{d\mathrm{E}}{d s_i^t} \simeq \frac{\partial \mathrm{E}}{\partial s_i^t},
\label{EQ:e_prop_approximation}
\end{equation}
 
where the symbol $\partial$ indicates that only direct contributions of $s_i^t$ on $\mathrm{E}$ should be accounted for in the derivation, thus removing the influences of spike $s_i^t$ on subsequent network activity (i.e. the elicit of spike $s_j^{t_+}$ for $t_+ > t$). This approximation is mandatory for a biologically plausible learning rule, which must satisfy space-time locality. If we substitute the approximation \eqref{EQ:e_prop_approximation} into eq. \eqref{EQ:exact_error_gradient} (and use the explicit expression for the error $\mathrm{E}$ in \eqref{EQ:error_definition} and \eqref{EQ:output_definition}) we get:
 
\begin{equation}
     \frac{d \mathrm{E}}{d w_{ij}} \simeq \sum_t \frac{\partial \mathrm{E}}{\partial s_i^t} p_i^t \hat{e}_j^t = 2 \left(\beta_\star-1 \right) \sum_{t, k} \left[ \sum_{\tau > t} \mathsf{B}_{ik} \left(y_k^{\star \tau} - y_k^{\tau} \right) \beta_\star^{\tau - t} \right]  p_i^t e_j^t,
\end{equation}
 
where $\beta_\star = \exp \left(-\Delta t / \tau_\star \right)$. Again, the apparent issue of the sum on future events can be solved by an exchange in the summation indices, which yields:

\begin{equation}
    \frac{d \mathrm{E}}{d w_{ij}} \simeq 2 \left(\beta_\star - 1 \right) \sum_{t, k} \mathsf{B}_{ik} \left( y_k^{\star t} - y_{k}^t \right) \sum_{\tau < t} \beta_\star^{t - \tau} p_i^{\tau} e_j^{\tau}.
\label{EQ:error_gradient_final_form}
\end{equation}

The previous expression for the error gradient can then be used as a synaptic update rule to improve network performances. In our experiments we have introduced an additional approximation to \eqref{EQ:error_gradient_final_form} by neglecting the following temporal filter:

\begin{equation}
    \sum_{\tau < t} \beta_\star^{t - \tau} p_i^\tau e_j^\tau \simeq \beta_\star p_i^{t-1} e_j^{t-1}.
\end{equation}

These additional approximations justify the use of the $\simeq$ symbol in eq. (6) of the main text and yield our final expression for the synaptic update rule (where all constant factors are included in the definition of the learning rate $\eta$):

\begin{equation}
    \Delta w_{ij} \simeq \eta \sum_t \left[ \sum_k \mathsf{B}_{ik} \left( y_k^{t+1} - y_k^{t+1} \right) \right] p_i^t e_j^t.
\end{equation}

Indeed, we observe that this approximation does not significantly affect the learning.  It is straightforward to derive the learning rule for the readout weights:

\begin{equation}
    \Delta B_{kj}= -\eta_\mathrm{RO}\frac{d \mathrm{E}}{d  B_{kj}} =  \eta_\mathrm{RO} \sum_{t} \left( y_k^{\star t} - y_k^t \right) \bar s_j ^t.
    \label{ro_rule}
\end{equation}

\section{Dimensionality of solution space}

This section provides more details about the section \textit{Dimensionality of solution space} and Fig.2 of the main text.

\subsection{Store and recall of a 3D trajectory and training protocol}
\label{SEC:Store_and_recall_3D_trajectory}

To investigate the properties of the solution space, we decided to store and recall a 3D continuous trajectory. Given a target input $x_h^t$ ($h = 1,..I$, $t = 1,..T$), the network should reproduce the target output $y_k^{\star \ t}$ ($k = 1,..O$, $t = 1,..T$).
$y_k^{\star \ t}$ is a temporal pattern composed of $3$ independent continuous signals. Each target signal is specified as the superposition of the four frequencies $f \in \left\{1, 2, 3, 5 \right\}$ Hz with uniformly extracted random amplitude $A \in \left[0.5, 2.0 \right]$ and phase $\phi \in \left[0, 2 \pi \right]$.
In this case the input, referred to as clock signal, is defined as follows. For $t \in (0,0.2 \, T]$, $x_1^t = 1$ while the other units are zero. For $t \in (0.2  \, T,0.4 \, T]$, $x_2^t = 1$ while the other units are zero and so on. The general rule can be written as $x_k^t = 1$ if $t \in \left(0.2 (k-1)\, T,0.2 k \, T\right]$, $x_k^t = 0$ otherwise.
In order to train our recurrent spiking network, the first step is to compute the target pattern of spikes $s_i^{\star t}$ is computed as:
\begin{align}
&\hat s_i^{\star t} = \exp{ \left(- \frac{\Delta t}{\tau_s}\right)} \, \hat  s_i^{\star t-1}  +\left( 1 - \exp{ \left(-\frac{\Delta t}{\tau_s} \right) }  \right) \, s_i^{\star t} \\
&v_i^{t} = \exp{ \left(- \frac{\Delta t}{\tau_m}\right)} \, v_i^{t-1} + \left( 1 - \exp{ \left(-\frac{\Delta t}{\tau_m} \right) }  \right) \, \left( \sum_j  {w}_{ij} \hat  s_j^{\star t-1} + I_i^t + v_\mathrm{rest} \right) - \, w_\mathrm{res} s_i^{\star t-1}\\
&s_i^{\star t+1} = \Theta \left[  v_i^t - v^\mathrm{th}  \right]
\label{s_star}
\end{align}

where $I_i^t =I_i^{\mathrm{teach},t} +I_i^{\mathrm{in},t}  $.  $I_i^{\mathrm{teach},t} = \sum_h w_{ih}^\mathrm{teach} y_{h}^{\star t+1}$ and $I_i^{\mathrm{in},t} = \sum_h w_{ih}^\mathrm{in} x_{h}^{\star t}$. This term is only used to compute the target, and is not present during the test.
$w_{ih}^{\mathrm{in},t}$ and $w_{ik}^{\mathrm{teach},t}$ are random matrix whose elements are randomly drown from a Gaussian distribution with $\mathrm{zero}$ means and respective variances $\sigma_\mathrm{in}$ and $\sigma_\mathrm{teach}$.
All the parameters are reported in Table \ref{table1}.

The recurrent weights are learned with the following rule (eq.(14) of the main text)

\begin{equation}
    \Delta w_{ij} = \eta \sum_t \left[ \sum_k  {\mathsf{D}}_{ik} \left( \bar s_k^{\star t+1} - \bar s_k ^{t+1} \right) \right] p_i ^t e_j ^t,
    \label{rec_rule}
\end{equation}

while the learning rule for the readout weights is defined in eq.\eqref{ro_rule}.

\begin{table}
  \caption{Parameters for the Store \& Recall}
  \centering
  \begin{tabular}{lll}
    \toprule
    \multicolumn{2}{c}{Network Parameters}                   \\
    \cmidrule(r){1-2}
    Name     & Description     & Value \\
    \midrule
    $N$ & Number of units  & $100$     \\
    $T$ & Expert demonstration duration & $100$     \\
    $I$ & Number of input units   & $5$  \\
    $0$ & Number of output units & $3$ \\
    $dt$ & Time step & $1\mathrm{ms}$ \\
    $\tau_m$ & Membrane time scale & $8\ dt$ \\
    $\tau_s$ & Synaptic integration time scale & $2\ dt$\\
    $\tau_\mathrm{RO}$ & Readout time scale & $20 \ dt$\\
    $\delta v$ & Pseudo-derivative width & $0.2$\\
    $v_\mathrm{rest}$ & Rest membrane potential & $-4$\\
    \midrule
    \\
    \multicolumn{2}{c}{Training Parameters}                   \\
    \cmidrule(r){1-2}
    $\sigma_J$ & Variance of initial weights & $0$\\
    $\sigma_\mathrm{in}$ & Variance of input matrix & $30$\\
    $\sigma_\mathrm{teach}$ & Variance of teach signal & $1.0$\\
    $\eta$ & Recurrent learning rate & $0.1$\\
    $\eta_\mathrm{RO}$ & Readout learning rate & $0.015$\\
    epochs & Training iterations & $1000$\\
    \bottomrule
  \end{tabular}
  \label{table1}
\end{table}

\subsection{PC representation and dimensionality estimation}

The learning dynamics can be easily pictured as a trajectory where a single point is a complete history of the network activity $ {s}_n =  \left\{s_i^t: i= 1,...N; t = 1,...T  \right\}$.
For simplicity, and for visualization purposes we defined each point of the trajectory as defined by the vector ${s}_n := \sum_t |s_i^{\star \ t}-s_i^t|$. 
Every $50$ learning steps a vector ${s}_n$ is collected. $10$ different realizations of the experiment are performed (for different initialization of the initial recurrent weights, which are randomly extracted from a Gaussian with zero mean and variance $2.0$).
This procedure provides $10$ different trajectories ${s}_n$. The first 2 PCs of these trajectories are reported in Fig.2A-B of the main text.

\subsection{Dimensionality estimation}

In order to estimate the dimensionality of the solution space we consider the difference between the activity generated by the network at the end of the learning procedure and the target sequence $ \delta s_i^t  = s_i^{\star \ t}-s_i^t$.

When the sequence is perfectly cloned, $ \delta s_i^t=0$ by definition. Otherwise these deviations are different from zero. In order to estimate the dimensionality of the sub-space of the solution  containing $\delta s_i^t$ we first perform the PC analysis. PCA is applied on a collection of $T = 100$ vectors $\delta s_i^t$ each of them defined by $N = 100$ coordinates. As a result we obtain $N$ principal component variances $\lambda_k$. If only one variance is significantly different from zero the dimensionality is approximately one, and so on. The dimensionality can be estimated as $d = \frac{(\sum_k \lambda_k )^2}{\sum_k \lambda_k^2}$.

\subsection{Random $\mathsf{R}$ matrix}

\begin{figure}
\centering
\includegraphics[width=.8 \textwidth]{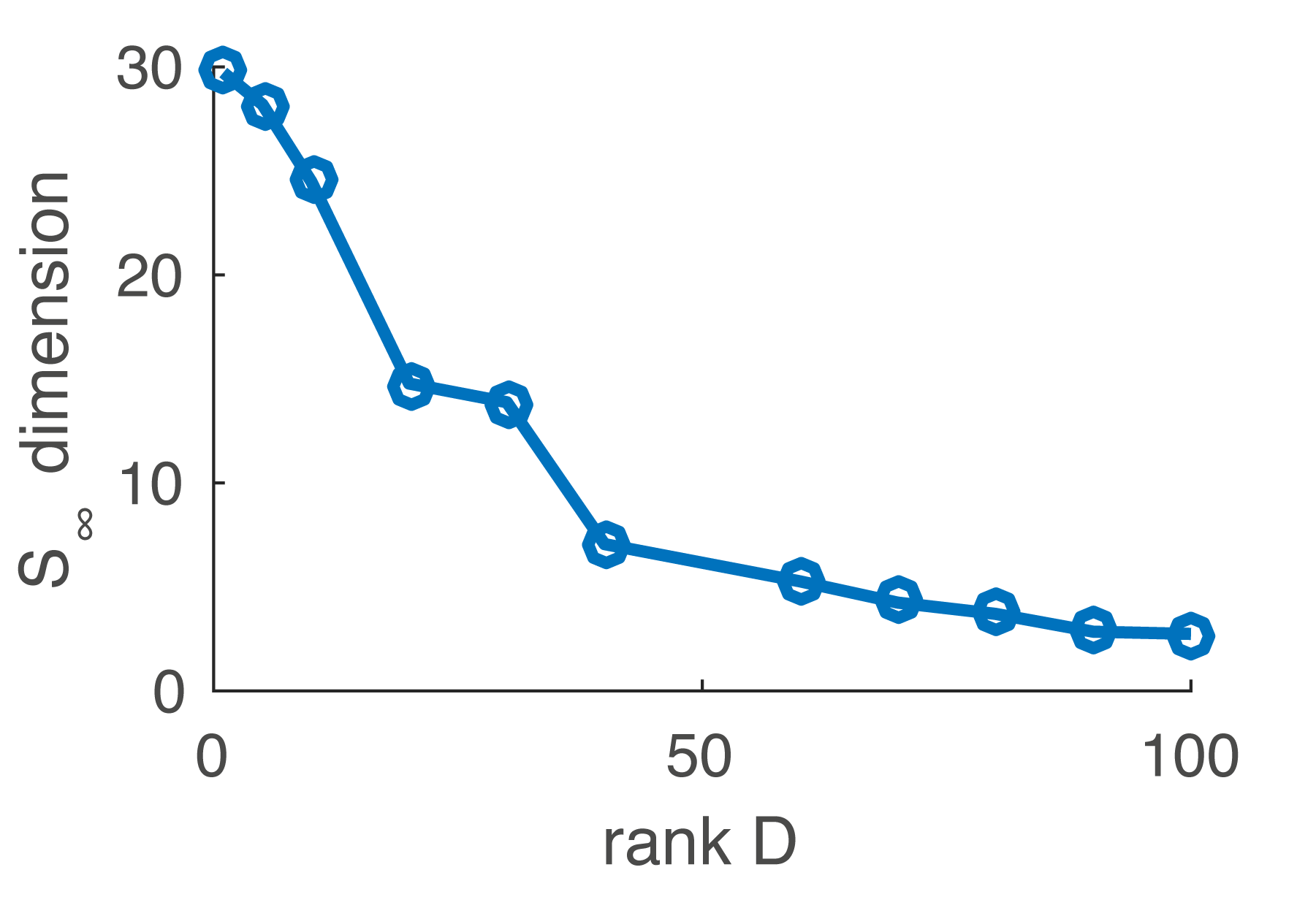}
\caption{ \textbf{Dimensionality of the solution space: $\mathsf{R}$ random.} Dimensionality of the solution space $\mathcal{S}_{\infty}$ as a function of the rank $\mathcal{D}$ of the error propagation matrix. }
\label{Fig_S1}
\end{figure}

We repeated the experiment of the main text (reported in Fig.2C) for the case in which the matrix $\mathsf{R}_{ik}$ is random. Its elements are randomly extracted from a Gaussian with mean zero and variance $\frac{1}{\sqrt{\mathcal{D}}}$. The result, reported in Fig.S\ref{Fig_S1} is analogous with what observed in the paper.

\section{Button-and-food}

\subsection{Training details}

\begin{figure}
\centering
\includegraphics[width=\textwidth]{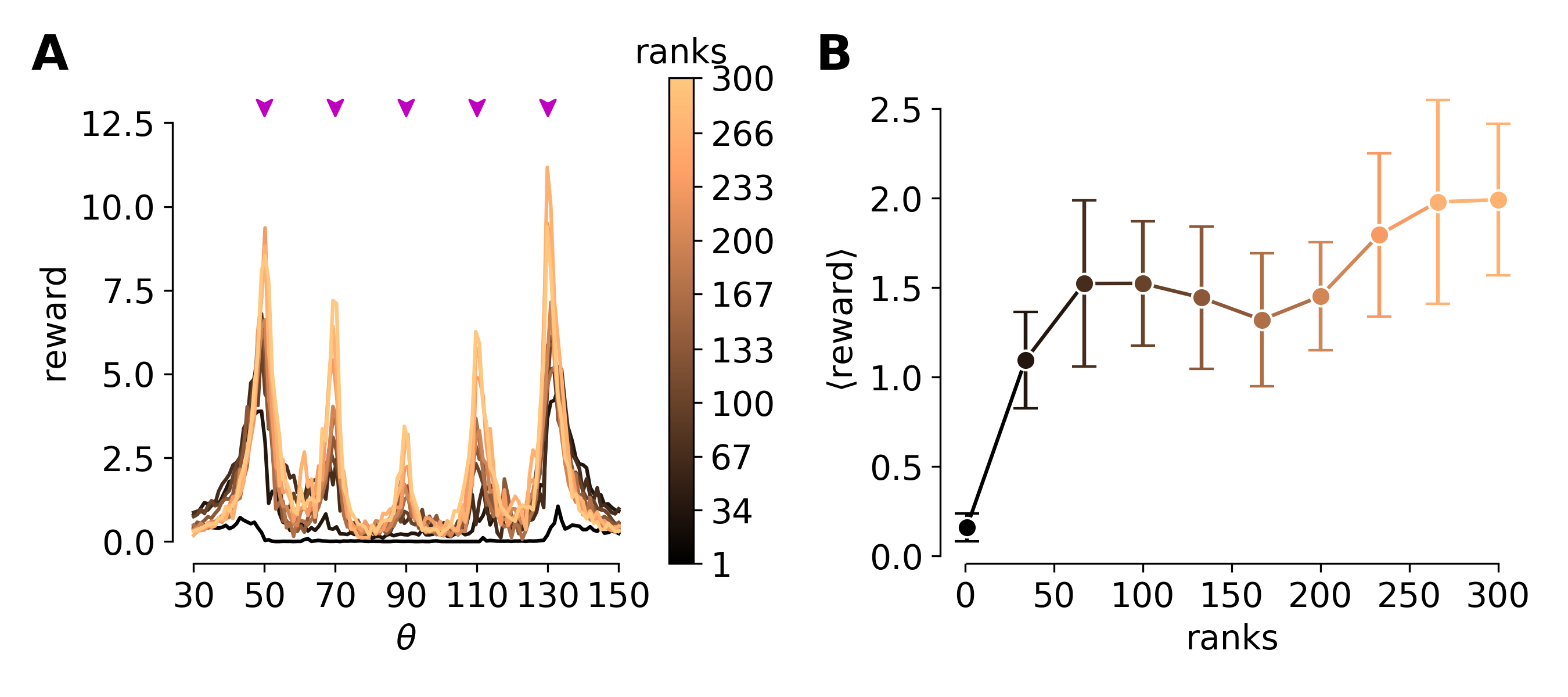}
\caption{\textbf{Button \& Food: $\mathsf{R}$ random.} (\textbf{A}) Average reward as a function of the test angle $\theta$ in the Button \& Food task. Purple arrows mark the training angles, while different colors code for different ranks of the feedback matrix $\mathsf{D}_{ik}$. (\textbf{B}) Average reward across all test angles as a function of the rank of the feedback matrix $\mathsf{D}_{ik}$. Error bars are standard deviations over approximately $100$ repetitions of the experiment. We used $N = 300$ and $T = 80$.}
\label{Fig_S2}
\end{figure}

The Button \& Food task requires an agent, which starts at the center of the environment domain, to reach for a button in order to unlock its final target (the food) which sits on a separate location. For this task we used the same learning procedure as described in Appendix \ref{SEC:Store_and_recall_3D_trajectory}. In particular, in the main text in Fig.4 of the main text we reported results obtained when training using the semi-clumped version of the learning rule. In the clumped version one substitutes the target spiking activity $s_i^{\star t}$ to the network activity $s_i^t$ during the evaluation of the spike response function $e_j^t$, indeed, upon convergence, learning halts precisely when $s_i^t = s_i^{\star t}$. However, when $\mathcal{D} < N$ and $\mathsf{D}_{ik}$ is diagonal, the target cannot effectively be enforced (or learned) for the $N - \mathcal{D}$ $\kappa$-neurons for which $\mathsf{D}_{\kappa \kappa} = 0$, so $s_i^{\star t}$ is only replaced to $s_i^{t}$ for $i \neq \kappa$, yielding the semi-clumped formulation.

In our experiments we set the initial agent position at coordinates $(x, y) = (0, 0)$, with the button positioned at $(0, -0.2)$. The target sits at a constant radius ($R_\star = 0.7$) from the origin, while the angle $\theta_\star$ is varied in the range $\left[30^\circ, 150^\circ \right]$. The agent is trained on the angles $\theta \in \left\{ 50, 70, 90, 110, 130 \right\}$ and tested on the complete range. We fix the network size and task temporal duration to $N = 300$ and $T = 80$ respectively. The expert trajectories are straight lines connecting agent initial location to button to food, travelled with constant speed so to match the travel time of $T^1_{a \to b} = 30$ and $T^2_{b \to f} = 50$. Additionally, a hint signal composed of two units ($H = 2$) encoding for the Boolean $\mathrm{locked}$ variable (square signal active when true) is injected on top of the expert trajectories (via a Gaussian random projection matrix of zero mean and variance $\sigma_\mathrm{hint}$) before computing the target activity.

The agent receives as input the difference of both button's and food's position with respect to current agent location, the input vector is then represented as $\Delta^t = \{ \Delta x^t_b, \Delta y_b^t, \Delta x_f^t, \Delta y_f^t \}$. This vector is then encoded via a set of tuning curves: the domain is partitioned in each direction ($x$ and $y$) with a resolution of $20$ cells, each cell coding a scalar input via a Gaussian activation centered around it physical location and of unit variance (with a peak value of $1$). Thus, each of the four input vector components $\Delta x^t_{b, f}, \Delta y_{b, f}^t$ are encoded using $20$ units, yielding a total input count of $I = 80$ units. The agent output is the velocity vector $v_{x, y}$, encoded using $O = 2$ output units. The agent reward is computed as $r_\mathrm{unlock} = \mathrm{min}\  d \left( p_\mathrm{agent}, p_\mathrm{target}  \right)^{-1}$ when the target is successfully unlocked (button pushed), while fixed at $r_\mathrm{locked} = 1 / d_\mathrm{max}$, $d_\mathrm{max} = 5$ for the locked condition. The feedback matrix $\mathsf{D}_{ik}$ is computed as $\mathsf{D} = \mathsf{R}^\top \mathsf{R}$, where $\mathsf{R}$ is a Gaussian random matrix of shape $\mathbb{R}^{\mathcal{D} \times N}$, with $\mathcal{D}$ the predefined matrix rank, and variance $\sigma = 1 / \sqrt{\mathcal{D}}$. The learning procedure is the same as described in Appendix B.1. We trained our system for $1000$ epochs (with Adam optimizer) and repeat the experiment $100$ times for each rank. 

\subsection{Random $\mathsf{R}$ matrix}

We report here results obtained for the pure non-clumped version of the rule \eqref{rec_rule} and random feedback matrix $\mathsf{R}_{ik}$.
In Fig.S\ref{Fig_S2} we report the results of the analysis. In Fig.S\ref{Fig_S2}A the average reward (error bars omitted for visualization clarity) as a function of the test angle $\theta$ is plotted for the different ranks of the feedback matrix $\mathsf{D}$, highlighting the superior performances of the high-ranks for this particular task. This features is then confirmed in panel Fig.S\ref{Fig_S2}B, where the average reward (across all test conditions) is reported as a function of the rank $\mathcal{D}$ of the feedback matrix $\mathsf{D}$, again stressing how a high-rank feedback induces optimal performances for this task. This result confirms that what reported in the main text for a diagonal feedback matrix $\mathsf{R}_{ik}$ is indeed general and holds for a random $\mathsf{R}_{ik}$ as well. The complete set of parameters used in the Button \& Food task is reported in Table \ref{table2}.

\begin{table}
  \caption{Parameters for the Button \& Food task}
  \centering
  \begin{tabular}{lll}
    \toprule
    \multicolumn{2}{c}{Network Parameters}                   \\
    \cmidrule(r){1-2}
    Name     & Description     & Value \\
    \midrule
    $N$ & Number of units  & $300$     \\
    $T$ & Expert demonstration duration & $80$      \\
    $I$ & Number of input units   & $80$  \\
    $0$ & Number of output units & $2$ \\
    $H$ & Number of hint units  & $2$ \\
    $dt$ & Time step & $1\mathrm{ms}$ \\
    $\tau_m$ & Membrane time scale & $6\ dt$ \\
    $\tau_s$ & Synaptic integration time scale & $2\ dt$\\
    $\tau_\mathrm{RO}$ & Readout time scale & $10 \ dt$\\
    $\tau_\star$ & Target time scale & $5\ dt$\\
    $\delta v$ & Pseudo-derivative width & $0.01$\\
    $v_\mathrm{rest}$ & Rest membrane potential & $-4$\\
    \midrule
    \\
    \multicolumn{2}{c}{Training Parameters}                   \\
    \cmidrule(r){1-2}
    $\sigma_J$ & Variance of initial weights & $1 / \sqrt{N}$\\
    $\sigma_\mathrm{input}$ & Variance of input matrix & $5$\\
    $\sigma_\mathrm{teach}$ & Variance of teach signal & $3$\\
    $\sigma_\mathrm{hint}$ & Variance of hint signal & $5$\\
    $\eta$ & Recurrent learning rate & $0.01$\\
    $\eta_\mathrm{RO}$ & Readout learning rate & $0.01$\\
    $\sigma_\mathsf{D}$ & Variance of feedback matrix & $1 / \sqrt{\mathcal{D}}$\\
    epochs & Training iterations & $1000$\\
    \bottomrule
  \end{tabular}
  \label{table2}
\end{table}

\section{2D Bipedal Walker}

\subsection{Training details}

The 2D Bipedal Walker environment was, provided through the OpenAI gym (\url{https://gym.openai.com} \cite{1606.01540}, MIT License). 
The expert behavior is obtained by training a standard feed-forward network with PPO (proximal policy approximation \cite{schulman2017proximal}, in  particular we used the code provided in \cite{pytorch_minimal_ppo}, MIT License). The average reward performed by the expert is $\langle r \rangle_\mathrm{exp} \simeq 180$ while a random agent achieves $\langle r \rangle_\mathrm{rnd} \simeq -120$.
The sequence of states-actions is collected in the vectors $ {y}^{\star \ t}_k $, $k = 1,... \, O$, $ {x}^{\star \ t}_h$, $h = 1,... I$, $t = 1,...T$. 
The learning procedure is the same as described in Appendix B.1.
All the parameters of the experiment are reported in Table \ref{table3}.
For this experiment we chose the maximum rank ($\mathcal{D} = N$). In this case, if the matrix  $\mathsf{R}_{ij}$ is extracted randomly (e.g. Guassian with zero mean) the matrix $\mathsf{D}_{ij}$ is almost diagonal. For this reason we take $\mathsf{D}_{ij} = \delta_{ij}$. We evaluate the performances for different values of $\tau_{\star}$.
The learning is divided in 2 phases. In the first one, the recurrent weights are trained in order to reproduce the internal target dynamics $s_i^{\star t}$ for $500$ iterations using gradient ascent.
In the second phase of the training procedure, only the readout weights are trained using equation eq.\eqref{ro_rule}.

\begin{table}
  \caption{Parameters for the Bipedal Walker 2D}
  \centering
  \begin{tabular}{lll}
    \toprule
    \multicolumn{2}{c}{Network Parameters}                   \\
    \cmidrule(r){1-2}
    Name     & Description     & Value \\
    \midrule
    $N$ & Number of units  & $500$     \\
    $T$ & Expert demonstration duration & $400$     \\
    $I$ & Number of input units   & $15$  \\
    $0$ & Number of output units & $4$ \\
    $dt$ & Time step & $1\mathrm{ms}$ \\
    $\tau_m$ & Membrane time scale & $4\ dt$ \\
    $\tau_s$ & Synaptic integration time scale & $2\ dt$\\
    $\tau_\mathrm{RO}$ & Readout time scale & $2 \ dt$\\
    $\delta v$ & Pseudo-derivative width & $0.2$\\
    $v_\mathrm{rest}$ & Rest membrane potential & $-4$\\
    \midrule
    \\
    \multicolumn{2}{c}{Training Parameters}                   \\
    \cmidrule(r){1-2}
    $\sigma_J$ & Variance of initial weights & $0$\\
    $\sigma_\mathrm{in}$ & Variance of input matrix & $2.0$\\
    $\sigma_\mathrm{teach}$ & Variance of teach signal & $3.0$\\
    $\eta$ & Recurrent learning rate & $0.3$\\
    $\eta_\mathrm{RO}$ & Readout learning rate & $0.00375$\\
    epochs & Training iterations & $500$\\
    \bottomrule
  \end{tabular}
  \label{table3}
\end{table}

\begin{figure}
\centering
\includegraphics[width=.8\textwidth]{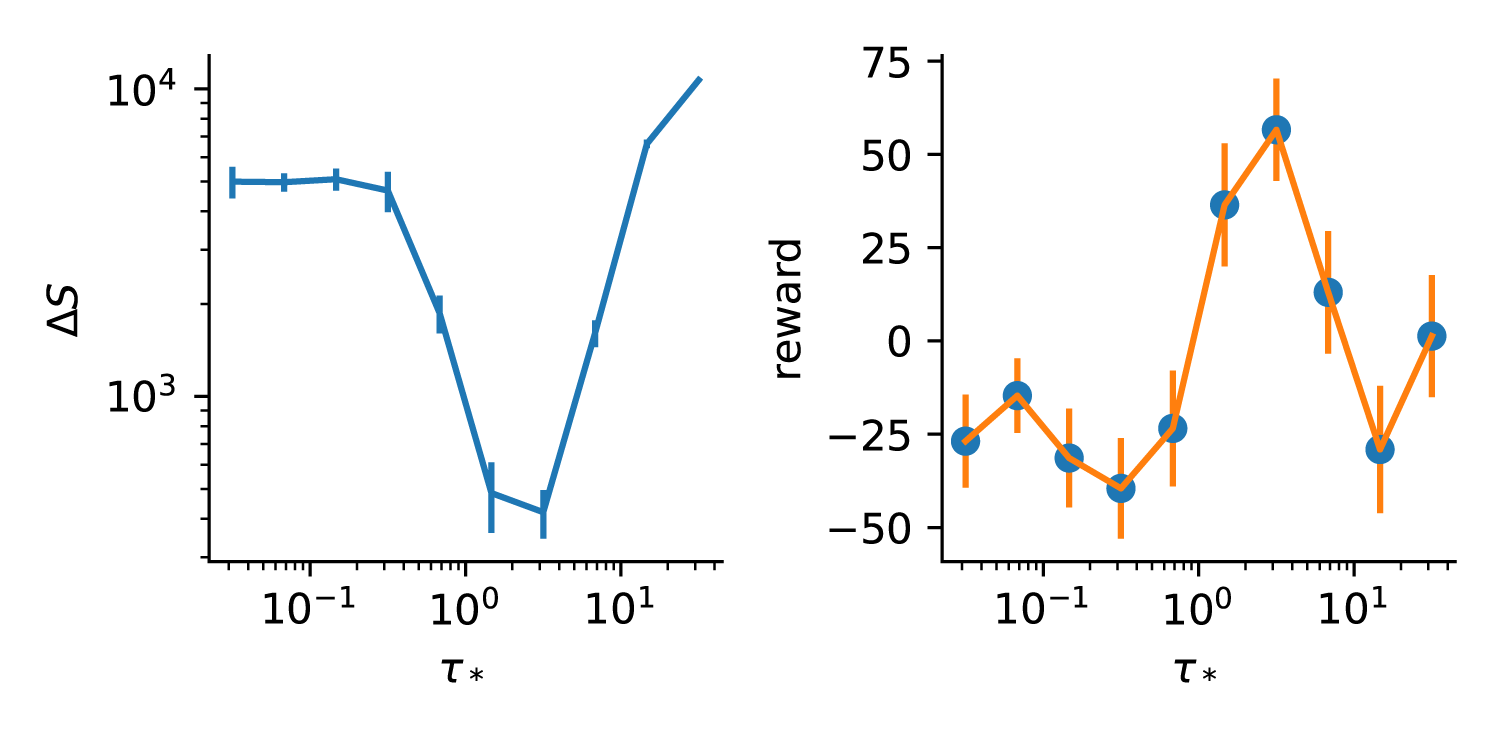}
\caption{\textbf{Bipedal Walker 2D. Non clumped version. } 
(left). Spike error $\Delta S$ as a function of the $\tau_{\star}$.
(right). Average reward as a function of the $\tau_{\star}$.
}
\label{Fig_S3}
\end{figure}

\subsection{Non-clumped case}

In the experiment reported in Fig.5 of the main text we adopted the clumped version of the learning rule. This means to substitute the target spiking activity $s_i^{\star t}$ to the activity produced by the network $s_i^{t}$ in the evaluation of the spike response function $e_j^t$.

We repeated the experiment in the non-clumped version of the training.
In Fig.S\ref{Fig_S3} it is reported the average of the maximum reward as a function of $\tau_{\star}$.
For each value of $\tau_{\star}$ we performed $10$ independent realizations of the experiment. For each realization the $s_i^{\star \ t}$ is computed, and the recurrent weights are trained. The optimization of the recurrent weights is performed using eq.\eqref{rec_rule} through gradient ascent and a learning rate $\eta$.
The learning of readout weights is performed using eq.\eqref{ro_rule} through gradient ascent and a learning rate $\eta_\mathrm{RO}$.
Every $75$ training iterations of the readout training we test the network and evaluate the average reward $\langle r \rangle$ over $50$ repetitions of the task. We then evaluate the average over the $10$ realizations of the maximum $\langle r \rangle$ obtained for each realization.

It is apparent that in this case there exist an optimal $\tau_{\star} \simeq 2.5 \mathrm{ms}$, allowing for a minimal training error ($\Delta S = \sum_{it}|s_i^{\star t}-s_i^{t}|$, the difference between the target pattern of spikes and the pattern generated is minimal,  Fig.S\ref{Fig_S3} left panel) and a maximum value of the reward ( Fig.S\ref{Fig_S3} right panel). However, the relationship between the training error and the reward is not trivial since for high $\tau_{\star}$ value we got a very poor training error, but also a good value of the average reward.